\documentclass[sigconf]{acmart}
\usepackage[ruled,vlined]{algorithm2e}
\usepackage{amsmath}

\DeclareMathAlphabet{\altmathcal}{OMS}{cmsy}{m}{n}
\usepackage{bm}
\usepackage{amsfonts}
\usepackage{enumitem}

\usepackage{multirow}
\usepackage{makecell}
\usepackage{arydshln}
\newcommand{\methodshort}{\textsc{DTDA}}
\newcommand{\method}{\textsc{Dynamic Threat Detection Agent}}

\newcommand{\hide}[1]{}

\AtBeginDocument{%
  }

\setcopyright{acmlicensed}
\copyrightyear{2026}
\acmYear{2026}
\acmDOI{XXXXXXX.XXXXXXX}
\acmConference[arXiv '26]{arXiv}{2026}{Washington, DC, USA}
\acmISBN{978-1-4503-XXXX-X/2018/06}

\begin{document}

\title{GenAI-Driven Threat Detection with Microsoft Security Copilot}

\author{Scott Freitas}
\authornote{Both authors contributed equally to this work.}
\affiliation{%
  \institution{Microsoft Security Research}
  \state{Arizona}
  \country{USA}}
\email{scottfreitas@microsoft.com}

\author{Amir Gharib}
\authornotemark[1]
\affiliation{%
  \institution{Microsoft Security Research}
  \city{Toronto}
  \country{Canada}
}
\email{agharib@microsoft.com}

\renewcommand{\shortauthors}{Scott Freitas \& Amir Gharib}

\begin{abstract}
Defending against today’s increasingly sophisticated cyberattacks requires security analysts to continuously translate evolving attacker tradecraft into detection logic. This places defenders in a reactive posture, requiring constantly updated expertise across an increasingly fragmented security landscape. We introduce the \method{} (\methodshort{}), an always-on adaptive agent that continuously investigates security incidents across Microsoft Defender to uncover hidden threats and generate explainable detections when attack-story gaps are found. \methodshort{} combines: (1) a unified activity timeline spanning alerts, events, user and entity behavior analytics, and threat intelligence; (2) versioned LLM prompt contracts with schema validation, grounding requirements, bounded retries, and fail-closed suppression; (3) a planner-executor investigation loop that generates attack-specific hypotheses and gathers supporting and refuting evidence; and (4) dynamic alert generation with a context-relevant title, severity, MITRE mappings, remediation guidance, implicated entities, and natural-language attack description.
Integrated into Microsoft Security Copilot and deployed across tens of thousands of Defender customers, \methodshort{} operates continuously at industry scale. In a 120-day online evaluation, \methodshort{} achieves 80.1\% precision from customer feedback while generating novel alerts for approximately 15\% of investigated incidents.
In offline evaluation, \methodshort{} recovers hidden malicious activity with 0.78 F1 using GPT-5.4, improving over GPT-4.1 by 0.12 F1 and outperforming the baseline by 0.26 F1 points.
Operationally, \methodshort{} processes single-incident investigations end-to-end in a median of 28 minutes at a median token cost of USD 2.04, with a 0.38\% job-level failure rate. These results demonstrate that autonomous agents can identify missed malicious activity at production scale while remaining practical for real security workflows.
\end{abstract}



\maketitle

\section{Introduction}
Enterprise security has advanced substantially in recent years, yet a fundamental gap remains: adversary behavior evolves continuously across complex enterprise environments, while static detection rules and supervised ML systems struggle to keep pace, often surfacing only partial views of malicious activity. Anomaly detection can help identify unusual behavior, but often produces noisy signals that require substantial analyst effort to validate.
In 2024, the average cost of a data breach reached \$4.4 million~\cite{ibm2025cost}, underscoring the cost of missed detection. At the same time, SOCs face a persistent scale mismatch: based on internal measurements, organizations can see up to 232 incidents per day, 51\% of which are never resolved. This makes exhaustive human investigation economically and operationally infeasible.

These pressures motivate investigation-driven detection, where incidents are autonomously examined beyond their existing alerts to uncover malicious activity.
Unified security operations platforms, such as Microsoft Defender, are uniquely positioned to enable this shift as they consolidate telemetry across security products into a cohesive view of the enterprise landscape~\cite{einav2023introducing,freitas2024graphweaver}. This broad visibility provides the context needed to identify malicious behavior that is often missed when products are analyzed in isolation. 
Microsoft Security Copilot builds on this foundation with a GenAI layer capable of reasoning over security incidents, creating an opportunity to move beyond analyst-assistive workflows toward continuous, autonomous threat discovery.

\vspace{1mm}\noindent \textbf{Autonomous threat detection.} Building industry-scale autonomous threat detection systems presents four fundamental challenges:

\begin{enumerate}[topsep=4pt, leftmargin=*, itemsep=3pt]
\item \textbf{Evidence fusion over fragmented telemetry.} Security evidence is distributed across hundreds of products and heterogeneous telemetry sources, including alerts, events, anomalies, and threat intel. An autonomous system must connect these fragmented sources into a coherent investigative context without losing critical detail or introducing excessive noise.

\item \textbf{High-precision under incomplete evidence.} Autonomous detections must meet the precision requirements of real security workflows. This requires reasoning over noisy, incomplete, and sometimes conflicting evidence, ruling out benign alternatives, and producing high-confidence alerts that analysts can trust.

\item \textbf{Open-world threat discovery.} Adversaries continuously evolve their tactics, requiring an autonomous system to uncover malicious behavior in an open-ended threat landscape rather than simply recognize pre-defined patterns.

\item \textbf{High-volume continuous investigation.} Delivering always-on protection requires investigating large numbers of incidents across massive telemetry volumes with low latency and sustainable cost, demanding a scalable and efficient architecture.
\end{enumerate}

\begin{figure*}
    \centering
    \includegraphics[width=\textwidth]{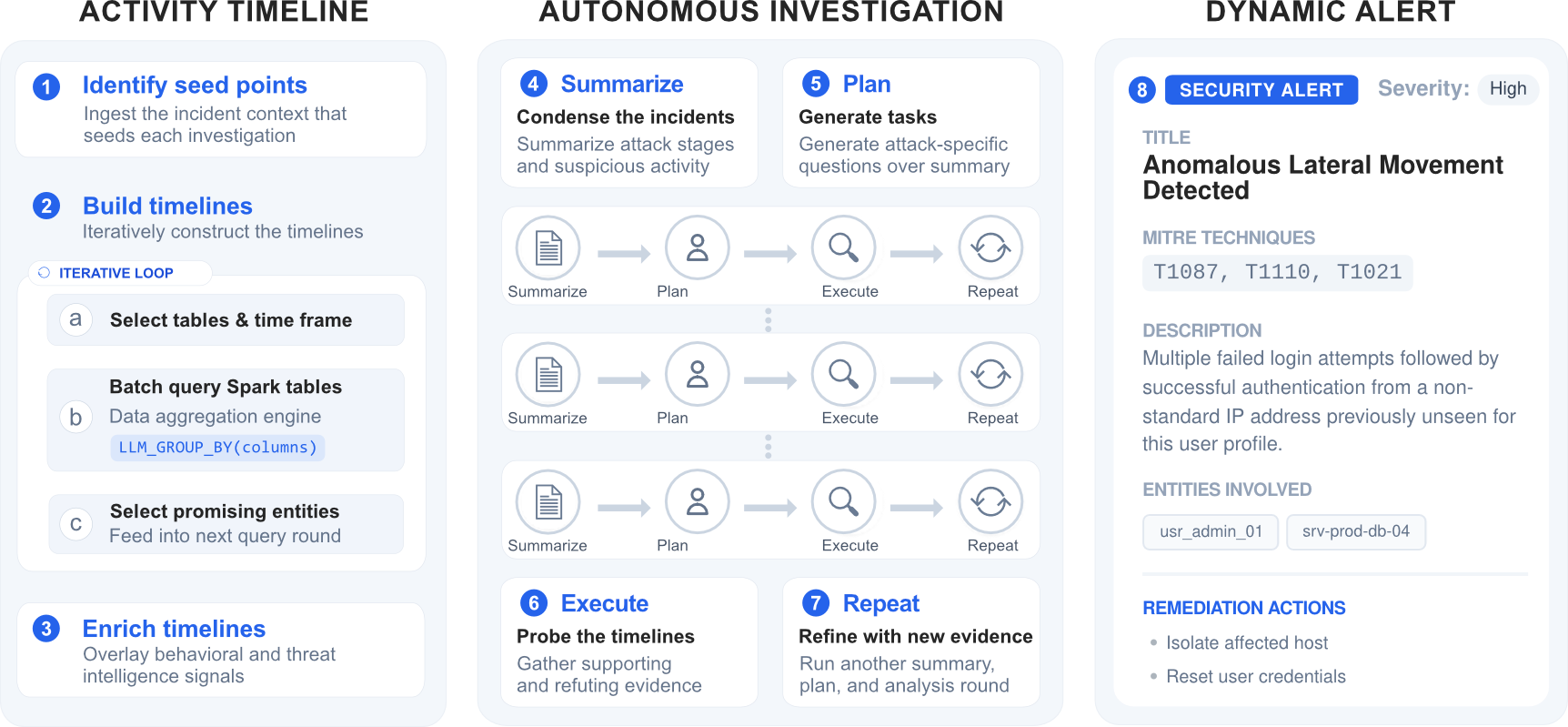}
    \caption{ Overview of the \methodshort{} architecture: an industry-scale framework for autonomous threat detection. \methodshort{} builds incident-centered activity timelines from alerts, events, UEBA, and threat intel; runs a bounded planner-executor investigation to gather supporting and refuting evidence; and emits a dynamic alert when the investigation identifies novel malicious activity.
    }
    \label{fig:crown}
\end{figure*}

\subsection{Contributions}
We introduce \methodshort{} (Figure~\ref{fig:crown}), a framework designed to address the challenges of autonomous threat detection at scale. Our framework makes significant contributions in the following areas:

\begin{itemize}[topsep=2mm, itemsep=0mm, parsep=1mm, leftmargin=*]
    \item \textbf{\methodshort{} architecture.} \methodshort{} advances the cybersecurity industry’s approach to autonomous threat detection through an architecture designed in close collaboration with security research experts. Key innovations include: (1) activity timeline construction through dynamic table selection, entity pivoting, and event aggregation; 
    (2) a planner-executor investigation loop that generates attack-specific hypotheses and gathers supporting and refuting evidence; and
    (3) gap assessment that determines whether the investigation uncovered novel malicious activity.
    By disclosing key architectural design elements and operational processes, we provide the first detailed public description of a production-scale GenAI-driven autonomous threat detection system.

    \item \textbf{Extensive evaluation.} We conduct a comprehensive evaluation of \methodshort{} across online customer feedback, controlled offline analysis, and production latency-cost measurement. In a 120-day online sample across 12 production regions, \methodshort{} received 1,088 alert-level customer grades from 208 organizations, achieving 80.1\% precision across initial access, execution, and post-compromise categories. For jobs processing a single incident, median end-to-end latency is 28 minutes, with a median token cost of USD 2.04 per investigation. In a controlled offline evaluation, \methodshort{} recovers hidden malicious activity with 0.78 F1 using GPT-5.4, improving over GPT-4.1 by 0.12 F1 points and outperforming the baseline by 0.26 F1 points.

    \item \textbf{Impact to Microsoft customers and beyond.} \methodshort{} is integrated into Microsoft Security Copilot and deployed across tens of thousands of organizations worldwide. 
    The introduction of \methodshort{} advances Microsoft Defender from analyst-assistive workflows toward continuous, autonomous threat discovery, with explainable Copilot-sourced alerts directly within workflows. 
\end{itemize}
\section{Background}
We review four areas of related work that inform \methodshort{}.

\subsection{Incident Correlation}\label{subsec:correlation_related}
Incident correlation links related alerts into cohesive incident narratives~\cite{freitas2024graphweaver,wu2019alert,elshoush2013intrusion,granadillo2016new,kotenko2022systematic}, providing the substrate for downstream analysis. 
In contrast, \methodshort{} starts from the correlated incident and expands beyond its alert set into adjacent events, user and entity behavior analytics (UEBA), and threat intelligence (TI) to construct a broader timeline for autonomous investigation.

\subsection{Provenance-Based Threat Investigation}\label{subsec:provenance_related}
Provenance-based security systems reconstruct attacks from audit logs by modeling causal relationships among processes, files, sockets, and other host-level entities~\cite{hossain2017sleuth,milajerdi2019holmes,freitas2020d2m, hassan2019nodoze,hassan2020tactical,milajerdi2019poirot,han2020unicorn}. These systems automate important parts of threat detection, triage, and hunting over noisy telemetry, but primarily operate deterministically over endpoint or host-level provenance graphs. In contrast, \methodshort{} investigates across a heterogeneous enterprise environment using a GenAI planner-executor loop to gather evidence, assess detection gaps, and emit dynamic alerts into security workflows.

\subsection{Guided Response}\label{subsec:gr_related}
Guided response systems assist analysts by recommending incident investigation steps, triage verdicts, or remediation actions~\cite{zhong2018cyber,jiang2024xpert,franco2020secbot,veeramachaneni2016ai,oliver2024carbon,alturkistani2022optimizing,foo2005adepts}. Copilot Guided Response (CGR)~\cite{freitas2025ai} is an industry-scale example of this process. In contrast, \methodshort{} runs continuously in the backend to autonomously uncover malicious activity.

\subsection{Agentic Security Investigation}\label{subsec:agent_related}
Recent research has explored GenAI agents for security investigation, including agentic alert-investigation workflows with constrained tool access over logs and text search~\cite{eilertsen2025towards}, multi-agent architectures for alert triage~\cite{wei2025cortex}, SOC assistants that combine planning, log retrieval, and enrichment tools~\cite{banstola2026socai}, and broader surveys and benchmarks for agentic AI in cybersecurity~\cite{lazer2026survey,wu2025excytin}. Industry systems also show growing momentum toward agentic security operations~\cite{google2026agentic,google2026use,palo2026cortex}. However, public industry documentation on autonomous threat detection typically emphasizes product capabilities without disclosing core mechanisms. 
Recent work also applies LLMs to vulnerability assessment, including code analysis, exploitability reasoning, and patch guidance~\cite{wang2026cybergym}. 
\methodshort{} addresses a different setting: rather than reasoning over software artifacts to find weaknesses, it reasons over live incidents and heterogeneous enterprise telemetry to uncover malicious activity.
\section{Architecture Overview}\label{sec:architecture}
\methodshort{}’s architecture and design choices were developed in close collaboration with security researchers to shape timeline construction and autonomous investigation around strategies used in human incident analysis.
The system operates as an always-on backend agent executed by regional jobs that run every few minutes to pick up prioritized security incidents for investigation.
We use PySpark’s distributed execution engine for telemetry retrieval, while reserving Python for last-mile timeline materialization and agentic reasoning steps that do not have native PySpark support.
\methodshort{} uses a stateless LLM endpoint: each call depends only on the prompt contract and explicitly provided incident context, telemetry summaries, or timeline evidence for that stage. This improves grounding and stability across parallel execution.
As illustrated in Algorithm~\ref{alg:pipeline}, \methodshort{} consists of two main components: 

\begin{itemize}[topsep=2mm, itemsep=0mm, parsep=1mm, leftmargin=*]
    \item \textbf{Activity timeline construction (Section~\ref{sec:timeline}).} Constructs activity timelines by combining alerts, events, UEBA, and threat intel into compact evidence substrates created through dynamic table selection, entity pivoting, and adaptive aggregation.

    \item \textbf{Autonomous investigation (Section~\ref{sec:investigation}).} Operating over the timeline, we develop a bounded planner-executor loop to generate attack-specific questions, gather supporting and refuting evidence, and assess whether the recovered evidence represents malicious activity not captured by the original incident alerts. 
\end{itemize}

\subsection{LLM Prompt Contracts}\label{subsec:prompt_contracts}
\methodshort{} uses LLM calls across table selection, aggregation planning, entity selection, planner-executor investigation, gap assessment, and dynamic alert generation. To reduce sensitivity to prompt wording and model changes, each LLM stage is implemented as a prompt contract with fixed input and output schemas, allowed values, grounding requirements, and deterministic validation checks. During production execution, invalid outputs are retried up to a bounded limit; repeated failures suppress the affected retrieval plan, task, evidence row, or candidate alert rather than allowing unsupported output to propagate downstream. This design makes prompt failures observable, limits drift across models, and allows \methodshort{} to degrade conservatively under malformed LLM outputs.

\subsection{Design Decisions and Trade-offs}\label{subsec:architecture_tradeoffs}
\methodshort{} is designed to recover evidence for missed malicious activity while remaining scalable. This creates several trade-offs:

\begin{itemize}[topsep=2mm, itemsep=0mm, parsep=1mm, leftmargin=*]
    \item \textbf{Timeline vs. graph.} \methodshort{} uses a timeline as the primary investigation substrate. While graphs model connectivity patterns, timelines preserve metadata in a form that is easy for the planner-executor loop to query and reason over.

    \item \textbf{Batched vs. interleaved.} \methodshort{} separates timeline construction from investigation.
    While a tightly coupled design could enhance investigation, it requires table query synchronization, increasing latency and keeping PySpark executors provisioned throughout the entire job. Batched retrieval instead amortizes queries across incidents, then releases executors once the timeline is complete. 
    
    \item \textbf{Gap discovery vs. triage.} \methodshort{} does not triage incidents; that responsibility is handled by other Security Copilot systems~\cite{freitas2025ai}. 
    \methodshort{} analyzes prioritized incidents~\cite{gharib2026introducing}, focusing on completing the attack story.
\end{itemize}

\section{Activity Timeline Construction}\label{sec:timeline}
Timeline construction is detailed across five key components.

\subsection{Incident Batching}\label{subsec:incident_seeding}
\methodshort{} continuously monitors security activity through PySpark jobs that run every few minutes and select incidents using predefined investigation criteria such as: priority level~\cite{gharib2026introducing}, critical asset involvement, disruption signals, and threat actors.
Selected incidents are grouped into batches by threat type, such as ransomware or initial access, to allow related investigations to reuse similar telemetry retrieval patterns to improve job efficiency.
For each incident, \methodshort{} retrieves the associated alerts and entities (e.g., users, devices, IPs) which define the initial investigative surface.

\subsection{Dynamic Table Selection}\label{subsec:dynamic_table_selection}
Microsoft Defender exposes a large and heterogeneous telemetry surface, so querying every possible source for every incident would be inefficient and noisy.
As a result, \methodshort{} uses an LLM-guided selection step to identify relevant tables.
Table selection is performed independently, but in parallel, allowing \methodshort{} to adapt the telemetry surface of each incident.
For each candidate table, the LLM receives structured incident context, the table schema, and a short description of the activity captured by the table.
Table selection is driven by the incident’s alerts, entities, and attack patterns.
The model returns a structured decision indicating whether each table should be queried and, if selected, the lookback window.

\subsection{Bounded Expansion}\label{subsec:entity_pivoting}
Using the selected tables and lookback windows,
\methodshort{} expands outward by querying telemetry linked to incident entities, summarizing high-volume events, and selecting entities for the next round. In practice, we find that two rounds of expansion provide a useful tradeoff between breadth and control: a single round often misses important follow-on activity, whereas multi-hop expansion can quickly introduce noise and cost.

\smallskip\noindent
\textbf{Expand.}
The selected telemetry tables are queried in parallel, with each query batched across incidents whose entities match the table schema to improve throughput.
Each retrieved row represents an event in which one or more incident entities appear alongside related entities and event metadata.
For example, an email row may connect a sender, recipient, and message identifier with a delivery status~\cite{microsoft2026understand}. 
This first expansion produces the neighborhood of events and entities for subsequent investigation.

\smallskip\noindent
\textbf{Aggregate.}
High-volume events from the expansion step are aggregated into higher-level activity summaries (Sec~\ref{subsec:event_aggregation}), while lower-volume events pass through unchanged. This reduces repetitive telemetry and avoids filling the context window, while preserving key entity relationships and semantics needed for investigation.

\smallskip\noindent
\textbf{Select.}
Next, \methodshort{} selects newly surfaced entities likely to advance the attack story. It first filters low-signal entities, such as common infrastructure, trusted services, or other unlikely security pivots, then uses an LLM-guided selection step to choose context-relevant entities for follow-on expansion.

\smallskip\noindent
\textbf{Repeat.}
\methodshort{} repeats table selection, expansion, and aggregation for one additional round using the selected carry-forward entities. The second-round retrieval plan may differ from the first because newly surfaced entities can reveal behaviors or attack stages that are connected to the original incident but not represented by its initial alerts. The resulting telemetry is merged with the first-round evidence to form a consolidated incident timeline.

\subsection{Adaptive Activity Summarization}\label{subsec:event_aggregation}
Raw telemetry is often too repetitive for downstream reasoning: even a scoped incident expansion can return thousands of low-level records that express the same investigative signal with minor variations. \methodshort{} compresses high-volume event patterns into activity summaries, lowering token usage and reasoning latency while preserving investigation-relevant behavior. Because useful aggregation keys vary by table schema and incident context, \methodshort{} does not rely on a fixed grouping template. Instead, for each table, an LLM receives the incident context, table description, pivot-entity distribution, and lightweight statistics over candidate grouping columns, such as null rate, distinct count, and largest-group concentration. The model returns a progressive grouping schedule with grouping keys, support thresholds, and rationale for each level.
Aggregation proceeds until the table falls under the row budget or the grouping limit is reached. \methodshort{} first applies the most detailed grouping level, aggregating only groups whose volume exceeds the support threshold while allowing lower-volume events to pass through unchanged. If the table remains too large, it applies progressively coarser grouping levels and repeats the process. For selected groups, \methodshort{} constructs hourly time-bin summaries containing event counts, entity counts, and sampled representative values.

\subsection{Behavioral and Threat Intel Enrichment}\label{subsec:timeline_enrichment}
After timeline construction, it is enriched with Microsoft UEBA signals~\cite{microsoft2026advanced}, which capture abnormal or high-risk behavior relative to established baselines. These signals provide context such as unusual access patterns, suspicious identity activity, or deviations from typical user, device, or resource behavior.
\methodshort{} then enriches entities with TITAN threat intelligence signals~\cite{freitas2025web}, adding reputational context across the broader security ecosystem. These signals are not treated as standalone verdicts; rather, they act as evidence multipliers that help strengthen or weaken activity interpretations.

\begin{algorithm}[!t]
\KwIn{Incidents $\mathbf{I}$, tables $\mathcal{T}$, UEBA $\mathcal{U}$, threat intel $\mathcal{K}$}
\KwOut{Dynamic alerts $\mathbf{D}$}

\SetKwBlock{Timeline}{Activity Timeline Construction}{end}
\SetKwBlock{Investigation}{Autonomous Investigation}{end}

\BlankLine
\Timeline{
$\mathbf{S} \leftarrow \text{IncidentBatching}(\mathbf{I})$ \\
$\mathbf{Q}_1 \leftarrow \text{SelectEntities}(\mathbf{S})$; $\mathbf{L} \leftarrow \mathbf{S}$ \

\For{$r \leftarrow 1$ \KwTo $2$}{
$\mathcal{P}_r \leftarrow \text{TableSelection}(\mathbf{S}, \mathbf{Q}_r, \mathcal{T})$ \\
$\mathbf{R}_r \leftarrow \text{Expand}(\mathbf{Q}_r, \mathcal{P}_r)$ \\
$\mathbf{X}_r \leftarrow \text{Aggregate}(\mathbf{R}_r, \mathbf{S})$ \\
$\mathbf{L} \leftarrow \mathbf{L} \cup \text{MaterializeRows}(\mathbf{X}_r)$ \\
$\mathbf{Q}_{r+1} \leftarrow \text{SelectEntities}(\mathbf{L})$ \\
}

$\mathbf{L} \leftarrow \text{EnrichTimeline}(\mathbf{L}, \mathbf{S}, \mathcal{U}, \mathcal{K})$ \
}

\BlankLine
\Investigation{
$\mathbf{E} \leftarrow \emptyset$ \\
$\mathbf{A} \leftarrow \text{SummarizeIncident}(\mathbf{S})$

\For{$b \leftarrow 1$ \KwTo $2$}{
$\mathbf{H}_b \leftarrow \text{Plan}(\mathbf{A}, \mathbf{L}, \mathbf{E})$ \\
$\mathbf{E} \leftarrow \mathbf{E} \cup \text{Execute}(\mathbf{H}_b, \mathbf{L})$ \\
}

$\mathbf{D} \leftarrow \text{GenerateAlerts}(\mathbf{A}, \mathbf{E})$ \\
}

\caption{\method{}}
\label{alg:pipeline}
\end{algorithm}
\section{Autonomous Investigation}\label{sec:investigation}
Unlike timeline construction, which expands the telemetry available for analysis, autonomous investigation operates over a fixed incident timeline to refine the interpretation of observed activity. For each incident, \methodshort{} runs an independent, bounded planner-executor loop that generates attack-specific questions, retrieves supporting and refuting evidence, and determines whether the recovered evidence reveals malicious activity not captured by the original incident alerts.

\subsection{Incident Summary}
\methodshort{} begins by constructing a structured summary of the alerts attached to the incident. Rather than processing each alert independently, the summary aggregates repeated alerts from the same detector, capturing the entities and metadata associated with each group. This compact representation preserves the original alert context and implicated entities without overwhelming the planner.

\subsection{Planner-Executor Loop}\label{subsec:planner_executor}
\begin{figure}
\centering
\includegraphics[width=\linewidth]{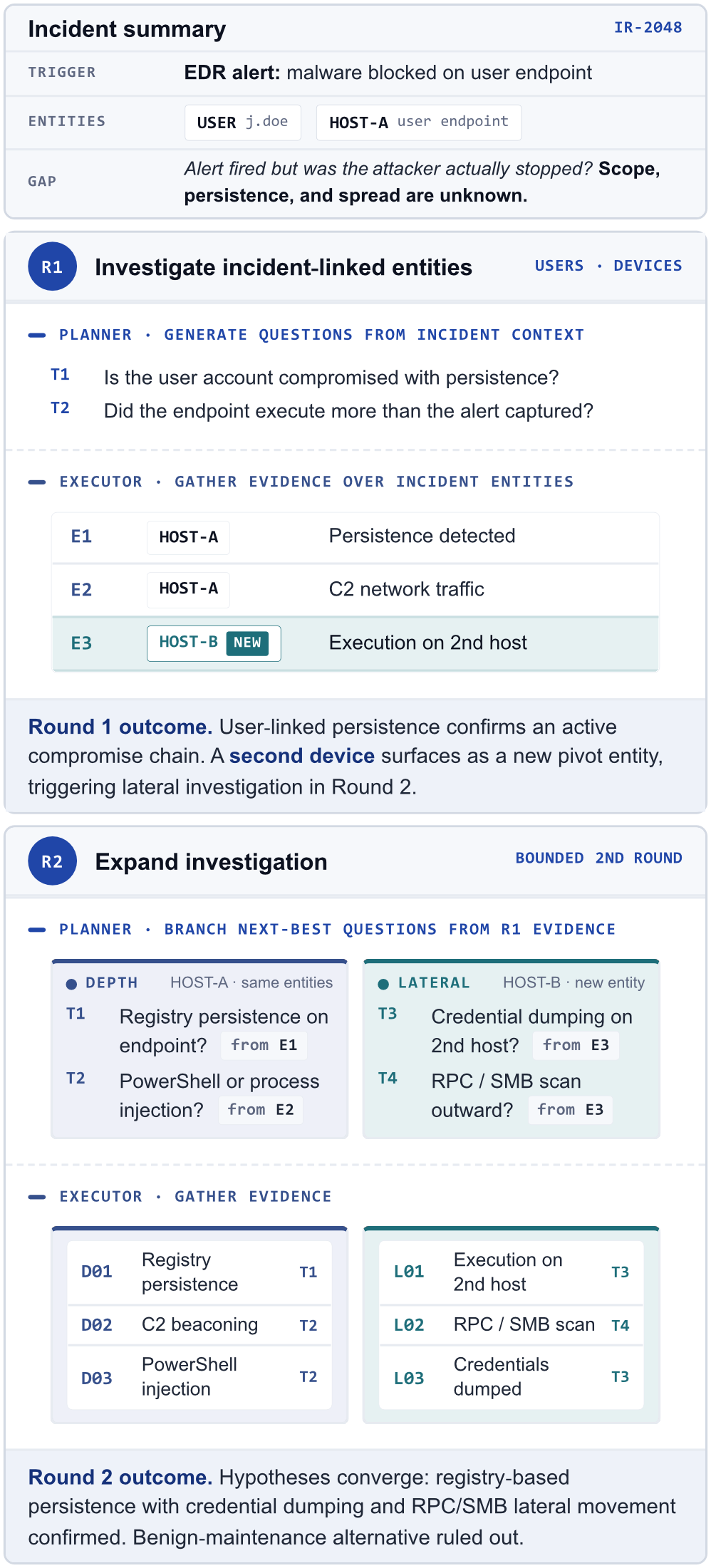}
\caption{Overview of the bounded planner-executor investigation loop. In Round 1, the planner generates targeted questions over incident-linked users and devices, and the executor gathers evidence that may surface new pivots. In Round 2, the system expands along bounded depth and lateral branches, using Round 1 evidence to focus follow-up questions, confirm attack progression, recover missing context, and rule out benign alternatives.
}
\label{fig:investigation_flow}
\end{figure}

The core reasoning engine is a bounded planner-executor loop operating over the incident summary and activity timeline (see Fig~\ref{fig:investigation_flow}). Rather than applying a fixed checklist, the loop iteratively asks attack-specific questions, retrieves relevant evidence, and updates the current interpretation of the incident.
To keep the investigation focused and aligned with timeline construction, \methodshort{} limits the planner-executor process to two rounds (R1 \& R2). This process operates under a budget that bounds the maximum number of investigative tasks based on the incident's prioritization score~\cite{gharib2026introducing}.

\medskip\noindent
\textbf{Planner.}
The planner converts the incident summary into a set of investigative tasks. It first forms candidate explanations for the observed activity, such as compromise, execution, persistence, lateral movement, or benign administrative behavior. Each task specifies the entity scope, hypothesis to test, and evidence needed to support, weaken, or redirect the current interpretation. In R1, the planner focuses on entities attached to the original incident alerts, reasoning over activity closest to the incident boundary. In R2, it branches into two task types based on what was found in R1: (1) \textit{depth tasks} that revisit incident-linked entities to validate or refute candidate explanations; and (2) \textit{lateral tasks} that investigate surfaced entities not part of the original incident.

\medskip\noindent
\textbf{Executor.}
The executor processes planner tasks in parallel. For each task, it gathers candidate evidence by matching the task’s entity values against timeline rows. 
Alert rows are excluded to avoid biasing the agent toward a malicious conclusion based only on existing alert associations; instead, the executor searches for independent behavioral and contextual evidence that supports or refutes the task hypothesis. Because entity matching can surface benign or weakly related activity, an LLM filters each candidate row in the context of the task and returns only relevant evidence, along with a brief explanation of how the row supports, refutes, or contextualizes the investigation.

\subsection{Dynamic Alerts}
\methodshort{} reviews the planner-executor analysis to determine whether the investigation uncovered an attack-story gap: malicious activity supported by timeline evidence but not represented by the existing incident alerts, such as a missing attack technique, or newly implicated entity. For each gap, the LLM generates candidate alerts with a title, description, severity, MITRE mappings, remediation guidance, implicated entities, and supporting evidence. Before emission, alerts must pass schema and grounding checks; alerts that lack evidence, duplicate existing activity, or fail validation are suppressed.
\section{Experiments}
To comply with privacy and data-residency requirements, \methodshort{} is replicated across geographic regions. We evaluate a sample of 12 production regions using online customer feedback from the Microsoft Defender portal and offline expert-labels.
Unless otherwise specified, online detection results are measured over a 120-day lookback using the GPT-4.1 model active during that window. Offline gap-recovery experiments are reported using GPT-4.1 model and GPT-5.4.
We report results using three attack-lifecycle phases: Initial Access (IA), Execution (EX), and Post Compromise (PC). This grouping provides a stable comparison 
while avoiding the sparsity of finer-grained categories in MITRE ATT\&CK~\cite{strom2018mitre}.

\medskip\noindent
\textbf{Limitations.}
We evaluate \methodshort{} as an industry-scale autonomous gap discovery system, rather than comparing against commercial alternatives whose architectures, telemetry, and outputs are undisclosed. 
Our offline held-out alert protocol provides a controlled measure of gap recovery, but cannot capture all forms of malicious activity, particularly attacks with no observable telemetry.

\subsection{Online Detection Quality}\label{subsec:online_detection_quality}

\textbf{Precision.}
Table~\ref{table:online_detection} reports alert-level precision from customer feedback, using labels assigned directly to \methodshort{} alerts rather than broader incident verdicts that may reflect other alerts in the same incident. Across 1,088 customer-graded alerts from 208 organizations, \methodshort{} achieves 80.1\% micro precision and 78.2\% macro precision. Precision is highest for post-compromise activity at 81.0\%, followed by execution at 80.7\% and initial access at 72.9\%. Initial access is the most challenging phase because early signals often have sparse context and can resemble benign anomalous authentication. Manual review shows that most false positives stem from ambiguous but grounded telemetry, benign administrative activity that resembles attack progression, or customer-feedback ambiguity. Execution and post-compromise alerts are stronger because they typically include richer corroborating evidence across the kill chain.

\medskip\noindent
\textbf{Representativeness.}
Because customer feedback is sparse and non-uniformly distributed, we measure organization-level coverage and concentration within the graded alert set. The 1,088 alert-level customer grades span 208 organizations, with the top-5 and top-10 organizations contributing 15\% and 23\% of graded \methodshort{} alerts, respectively. This indicates that the evaluation is not dominated by a small number of customers.

\medskip\noindent
\textbf{Novelty.}
\methodshort{} generated alerts in 15\% of investigated incidents.
To characterize this impact, Microsoft security experts manually reviewed a stratified sample of 1,764 alerts across attack phases and severity levels. Reviewers judged novelty relative to the original incident alert set using a fixed rubric: an alert was novel if it recovered a previously uncovered attack stage or technique, or implicated a new entity.
Because categories are multi-label, a single alert may satisfy multiple criteria. Of the reviewed alerts, 83.7\% recovered a missing attack stage, 88.8\% surfaced a newly 
implicated entity, and 1.5\% were judged redundant or low-value.

\subsection{Offline Gap Recovery}\label{subsec:offline_gap_recovery}
We complement the online analysis with a controlled held-out evaluation that measures whether \methodshort{} can recover malicious activity hidden from an incident. Because expert review is costly, we use a high-value cohort of 10 randomly sampled complex incidents involving multi-stage activity, such as ransomware, and reuse the same cohort across attack phases, model variants, and repeated runs. For each incident-phase example, we remove all alerts from the target phase and evaluate whether \methodshort{} recovers the missing activity from the remaining raw telemetry. Microsoft security experts assess recovery using a fixed rubric that compares each emitted alert and its supporting evidence against the held-out ground truth.

\medskip\noindent
\textbf{Model comparison and repeatability.}
To measure repeatability under LLM nondeterminism, each offline experiment is repeated 3 times on the same frozen snapshot. Table~\ref{table:offline_detection} reports performance by removed phase, with macro and micro averages across phases. We report precision and recall for GPT-5.4, and compare F1 against GPT-4.1. 
Overall, GPT-5.4 achieves 0.78 macro-F1 and 0.77 micro-F1, improving over GPT-4.1 by 0.12 and 0.11 F1 points, respectively. These results show that \methodshort{} can recover malicious activity hidden from the incident while remaining stable across repeated runs, and that the newer reasoning model improves gap-recovery quality.

\medskip\noindent
\textbf{Baseline.}
To isolate the value of structured investigation, we compare full \methodshort{} against a row-only variant that replaces the planner-executor loop with an LLM classifier, independently judging whether each timeline row reveals malicious activity not represented by the incident alerts.
With GPT-5.4, \methodshort{} improves macro-F1 from 0.52 to 0.78 and micro-F1 from 0.50 to 0.77 over the baseline. These gains reflect improvements in both precision and recall: macro precision increases from 0.64 to 0.97, while macro recall increases from 0.44 to 0.66.
Improvements are consistent across attack phases, with F1 increasing by 0.20 for IA, 0.33 for EX, and 0.24 for PC.
This shows that the planner-executor loop improves gap recovery beyond independent row-level classification, reducing false positives while recovering more malicious activity.

\begin{table}[t]
 \centering
 \setlength{\tabcolsep}{3pt}
 \begin{tabular*}{\linewidth}{@{\extracolsep{\fill}}lrrrrr@{}}
 \toprule
 \textbf{Phase} & \textbf{\# TPs} & \textbf{\# FPs} & \textbf{\# Orgs} & \textbf{Pr} & \textbf{95\% CI} \\
 \midrule
 IA & 86  & 32  & 71  & 72.9\% & 64.1--80.2\% \\
 EX & 159 & 38  & 78  & 80.7\% & 74.6--85.7\% \\
 PC & 626 & 147 & 150 & 81.0\% & 78.1--83.6\% \\
 \addlinespace[3pt]
 \cdashline{1-6}
 \addlinespace[3pt]
 \textbf{Micro} & 871 & 217 & 208 & 80.1\% & 77.6--82.3\% \\
 \textbf{Macro} & --  & --  & --  & 78.2\% & 74.8--81.6\% \\
 \bottomrule
 \end{tabular*}
 \caption{Online alert-level precision by attack phase across customer-graded \methodshort{} alerts using GPT-4.1.
 Labels reflect feedback on the emitted alert itself, not the broader incident. We report Wilson 95\% CI for phase-level and micro precision.}
 \label{table:online_detection}
 \vspace{-3mm}
\end{table}
 \begin{table}[t]
 \centering
 \setlength{\tabcolsep}{2pt}
 \begin{tabular*}{\linewidth}{@{\extracolsep{\fill}}lrrrrrrr@{}}
 \toprule
 \textbf{Phase} & $\mathbf{\bar{A}_{rm}}$ & $\mathbf{\bar{A}_{vis}}$ & \textbf{Pr} & \textbf{Re} & $\mathbf{F1_{4.1}}$ & $\mathbf{F1_{5.4}}$ & $\mathbf{\Delta F1}$ \\
 \midrule
 IA & 2.3 & 10.3 & .95 & .70 & .76$\pm$.03 & .80$\pm$.09 & +.04 \\
 EX & 4.8 & 8.9  & .99 & .59 & .65$\pm$.03 & .73$\pm$.10 & +.08 \\
 PC & 3.0 & 10.9 & .97 & .68 & .54$\pm$.15 & .80$\pm$.06 & +.26 \\
 \addlinespace[2pt]
 \cdashline{1-8}
 \addlinespace[2pt]
 \textbf{Macro} & 3.4 & 10.0 & .97 & .66 & .65$\pm$.07 & .78$\pm$.08 & +.12 \\
 \textbf{Micro} & 3.3 & 10.0 & .97 & .64 & .66$\pm$.03 & .77$\pm$.03 & +.11 \\
 \bottomrule
 \end{tabular*}
 \caption{Offline gap-recovery performance by attack phase. $\bar{A}_{rm}$ and $\bar{A}_{vis}$  denote the average number of alerts removed and visible per incident, respectively.
 Precision and recall are reported for GPT-5.4, while F1 is reported as mean $\pm$ std across $3$ repeated runs for GPT-4.1 and GPT-5.4.}
 \label{table:offline_detection}
 \vspace{-7mm}
 \end{table}

\subsection{Scale, Latency, Cost, and Stability}\label{subsec:latency_and_cost}
Operational metrics are collected from production logs at different granularities depending on the instrumentation available. For each metric, we state the denominator explicitly.

\medskip\noindent
\textbf{Scale.}
Across a sample of 10.3k incidents, \methodshort{} selected a median of 6 telemetry tables per incident and retrieved a median of 3,456 raw rows before aggregation. Adaptive aggregation reduced this evidence to a median of 887 timeline rows per incident, corresponding to a 2.4$\times$ median compression ratio. The distribution is heavy-tailed: average pre- and post-aggregation row counts were 68,447 and 8,548, respectively, with a 9.4$\times$ average per-incident compression ratio. This compression keeps LLM runtime and cost bounded while preserving key evidence for gap recovery.

\begin{figure}
    \centering
    \includegraphics[width=\linewidth]{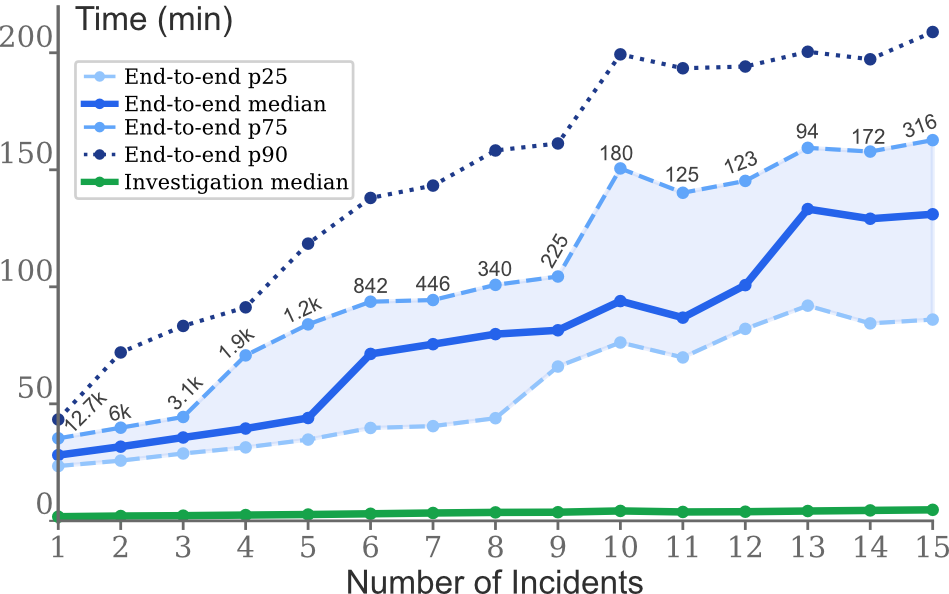}
    \caption{Runtime scaling by number of incidents per job, line labels indicate the number of runs. 
    Blue line shows median job latency, the shaded region shows p25--p75, and the top line shows p90. 
    Green line shows median investigation time.}
    \label{fig:runtime_scaling}
    \vspace{-3mm}
\end{figure}

\medskip\noindent
\textbf{Latency.}
Across 28k jobs, Figure~\ref{fig:runtime_scaling} shows that end-to-end latency increases predictably with the number of incidents per job, while autonomous investigation time remains nearly flat due to parallel incident reasoning. Timeline construction accounts for 89\% of end-to-end latency on average, making batched telemetry retrieval and aggregation the dominant runtime bottleneck and the primary target for future optimization. 
Single-incident jobs complete in a median of 28 minutes, and jobs can scale horizontally through increased concurrency subject to compute quota and cost controls.

\medskip\noindent
\textbf{Cost.}
We report token cost because it is directly attributable to each investigation, whereas compute and storage costs depend on regional capacity, workload concurrency, and internal infrastructure pricing. Across 90k incidents, \methodshort{} incurs a median token cost of USD 2.04 per investigation, with p75 and p95 costs of USD 3.36 and USD 7.82, respectively. Token cost is primarily driven by the size of the timeline and the number of tasks selected for analysis. These results show that the LLM reasoning component can operate at a practical cost for autonomous investigation.

\medskip\noindent
\textbf{Stability.}
\methodshort{} tracks structured-output validity and job-level failures in production. Schema-validation errors occur in 0.46\% of LLM responses and are retried when recoverable; repeatedly invalid outputs are suppressed rather than emitted. Across more than 236k production jobs, \methodshort{} had a 0.38\% job-level failure rate after retry exhaustion, showing stable production operation and conservative degradation under infrastructure or model-output failures.

\section{Deployment and Safety}\label{sec:deployment}

\textbf{Deployment.} 
\methodshort{} is replicated across geographic regions using Synapse. The production infrastructure has four main components: (a) ADLS for telemetry storage and access; (b) an Azure Synapse backend for deployment, orchestration, and monitoring of jobs; (c) elastic PySpark pools that scale active executors based on load; and (d) a production LLM endpoint accessed through Security Copilot. 
To support targeted retrieval over massive Parquet datasets, ADLS tables are optimized with partitioning, predicate pushdown, Bloom filters, Z-ordering, and column statistics.

\medskip\noindent
\textbf{Safety.} 
Because \methodshort{} operates as a backend service, users do not directly prompt the agent. Its inputs are product telemetry, which may be attacker-controlled. To mitigate prompt-injection and unsafe-output risks, \methodshort{} uses structured inputs, schema-validated outputs, and built-in Azure AI Content Safety mechanisms for harmful-content detection, prompt-injection and jailbreak mitigation~\cite{microsoft2026content}.
Together with the prompt-contract and grounding controls in Sec.~\ref{subsec:prompt_contracts}, these safeguards prevent unsupported free-form generations from being emitted into customer workflows.

\section{Future Research Directions}
Several directions can advance autonomous threat detection. First, future systems can explore scalable methods for interleaving investigation and timeline construction, allowing agents to gather evidence adaptively as hypotheses evolve. Second, continuous feedback loops can help systems adapt to changing environments by incorporating customer grades, analyst decisions, and emerging threat reports into investigation workflows. 
Third, security specialized fine tuning can improve core reasoning capabilities such as hypothesis generation, evidence selection, and benign explanation. Fourth, future systems must address adversarial adaptation, where attackers may shape telemetry to evade or confuse LLM-based reasoning. Finally, investigation findings can be translated into durable detection logic to close recurring gaps.
\section{Conclusion}
\methodshort{} demonstrates how GenAI can be deployed as part of the detection stack for modern security operations. By combining incident-centered timeline construction, bounded planner-executor investigation, and dynamic alerting, \methodshort{} moves beyond analyst-assistive workflows toward continuous autonomous discovery of malicious activity.
In production, \methodshort{} achieves 80.1\% alert-level precision from customer feedback and generates dynamic alerts for approximately 15\% of investigated incidents. In offline evaluation, \methodshort{} achieves 0.78 F1 with GPT-5.4, improving over GPT-4.1 by 0.12 F1 and outperforming a row-only gap detector by 0.26 F1 points. 
Operationally, \methodshort{} completes single-incident investigations in a median of 28 minutes at a median token cost of USD 2.04, while maintaining a 0.38\% job-level failure rate.
Integrated into Microsoft Security Copilot and deployed across tens of thousands of Defender organizations worldwide, \methodshort{} shows that autonomous investigation can uncover gaps at machine scale and surface explainable detections directly into security workflows.
\begin{acks}
We thank all of our colleagues who supported this research.
\end{acks}

\bibliographystyle{ACM-Reference-Format}
\bibliography{main.bib}

\end{document}